%% file: main.tex
\documentclass[10pt,conference]{IEEEtran}
\IEEEoverridecommandlockouts

\usepackage{quantikz}
\usepackage{dsfont}
\usepackage{braket}
\usepackage{comment}
\usepackage{cite}
\usepackage{amsmath,amssymb,amsfonts}
\usepackage{algorithmic}
\usepackage{graphicx}
\usepackage{textcomp}
\usepackage{xcolor}
\usepackage[inline,shortlabels]{enumitem}
\usepackage{caption}
\usepackage{subcaption}
\usepackage{pbalance}
\usepackage{xfrac}

\usepackage[
  colorlinks=true,
  linkcolor=blue,
  citecolor=blue,
  urlcolor=blue
]{hyperref}

\def\BibTeX{{\rm B\kern-.05em{\sc i\kern-.025em b}\kern-.08em
    T\kern-.1667em\lower.7ex\hbox{E}\kern-.125emX}}

\usepackage{xspace}
\newcommand{\brand}{COSMA\xspace}

\newlength{\mycolwidth}

\makeatletter
\if@twocolumn
\else
\fi
\makeatother

\begin{document}

\title{\brand: Communication-aware Optimization of Fermionic Simulation Kernels for Modular Quantum Architectures\thanks{This work has been accepted for presentation at the 2026 IEEE International Conference on Quantum Computing and Engineering (QCE). }}

\author
{\IEEEauthorblockN{Enrico Russo, Francesco G. Blanco, Elio Vinciguerra, Davide Patti, Giuseppe Ascia, Maurizio Palesi}
\IEEEauthorblockA{
Department of Electrical, Electronic, and Computer Engineering (DIEEI), University of Catania, I-95125 Catania, Italy\\ \textit{enrico.russo@unict.it, francesco.blanco@phd.unict.it, elio.vinciguerra@phd.unict.it, \{name.surname\}@unict.it}
}
}

\maketitle

\begin{abstract}
Quantum simulation is a leading application of quantum computing, but scaling to chemically relevant problems requires modular architectures composed of interconnected quantum processing units. In such systems, inter-core quantum communication becomes a major performance bottleneck. In this work, we present \brand, a communication-aware compilation framework for fermionic simulation kernels targeting modular quantum architectures. Our approach jointly optimizes fermion-to-qubit mapping, Pauli scheduling, and qubit allocation to minimize inter-core state transfers. Evaluated on molecular benchmarks, \brand achieves up to $2.5\times$ reduction in communication cost compared to state-of-the-art baselines, with a median improvement of $1.7\times$. These results demonstrate that cross-layer co-design is essential for efficient and scalable quantum simulation on multi-core quantum hardware.
\end{abstract}

    
\begin{figure*}[ht]
    \centering
    \includegraphics[width=\linewidth]{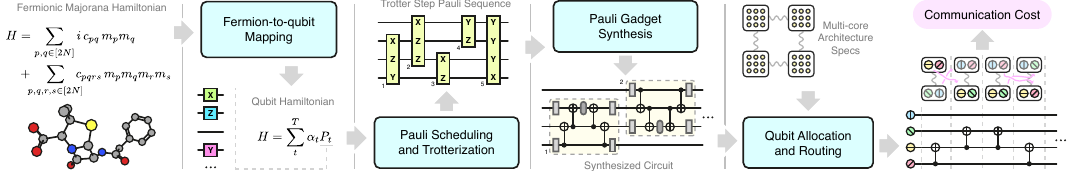}
    \caption{Overview of the compilation flow for quantum simulation kernels on modular architectures.}
    \label{fig:background}
\end{figure*}

\section{Introduction}

Quantum simulation is widely regarded as one of the most promising near- and long-term applications of quantum computing~\cite{chien2026simulating}. Since the seminal works of Manin and Feynman~\cite{manin2007mathematics,feynman2018simulating}, quantum computers have been expected to provide exponential advantages for modeling complex quantum systems that are intractable classically. In particular, quantum simulation is central to computational chemistry and materials science, enabling accurate modeling of strongly correlated electronic systems. These capabilities underpin a range of high-impact design workflows, including structure-based drug discovery, catalyst development for sustainable chemical transformations, and carbon-conversion technologies~\cite{cao2019quantum,sadybekov2023computational}.

Achieving practical quantum advantage in chemistry, however, requires a large number of logical qubits. Estimates suggest that moderately complex systems require tens to hundreds of logical qubits, while strongly correlated systems such as the FeMoco active site of nitrogenase may require on the order of $10^3$ logical qubits for accurate simulation~\cite{reiher2017elucidating}. These requirements are further amplified by quantum error correction (QEC), where each logical qubit is encoded into hundreds or thousands of physical qubits depending on target error rates~\cite{fowler2012surface}. As a result, large-scale fault-tolerant quantum computers may ultimately require millions of physical qubits.

Scaling to this regime poses significant architectural challenges. Most current devices adopt a monolithic design, where all qubits reside on a single processor with limited connectivity. As system size grows, such architectures face increasing issues in control complexity, cross-talk, fabrication yield, thermal management and wiring. Modular or multi-core quantum architectures have therefore been proposed, where multiple smaller quantum processing units (QPUs) or cores are interconnected via quantum communication links~\cite{monroe2014large,rodrigo2021double}. While this approach improves scalability and manufacturability, it introduces a critical bottleneck: inter-core quantum communication. Operations such as state transfer via teleportation or distributed entanglement~\cite{russo2025telesabre, caleffi2024distributed} are costly and can dominate execution time if not carefully minimized and can also introduce additional noise.

In this work, we address the problem of reducing inter-core communication for quantum simulation workloads on modular architectures. We propose \brand, a communication-aware compilation framework that jointly optimizes three key stages: (i) fermion-to-qubit mapping of the electronic Hamiltonian, (ii) scheduling of Pauli operators in the Trotterized time-evolution circuit, and (iii) co-optimization of Pauli gadget synthesis, qubit allocation, and routing.

By jointly optimizing these tightly coupled stages, \brand significantly reduces the number of inter-core quantum state transfers required during execution. This cross-layer co-design is particularly well suited to modular quantum computers, where communication costs dominate performance and scalability.

The main contributions of this work are:
\begin{enumerate}
    \item To the best of our knowledge, the first co-design study of fermion-to-qubit mapping, Pauli scheduling, and qubit allocation for quantum simulation kernels in modular architectures.
    \item A genetic algorithm for optimizing fermion-to-qubit mappings with respect to inter-core communication cost.
    \item A fast heuristic for parity tree synthesis and qubit allocation tailored to sequences of Pauli gadgets in modular quantum systems.
    \item An open-source, GPU-accelerated framework for evaluating mapping, scheduling, and allocation strategies for modular quantum architectures, available at \mbox{\url{https://github.com/haimrich/cosma}}.
\end{enumerate}

\input{background}

\section{Problem Formulation}
\label{sec:problem}

We formalise the optimization problem addressed by \brand and analyse the complexity of its constituent subproblems.

The input consists of a fermionic Hamiltonian $H$ with $N$ modes in Majorana form~\eqref{eq:hamiltonian_majorana} and a multi-core hardware graph $\mathcal{G}=(\mathcal{C},\mathcal{E})$, where each core $c\in\mathcal{C}$ has qubit capacity $\kappa$ and $d(c,c')$ denotes the shortest-path inter-core distance. Since PPTT mappings encode $N$ fermionic modes into exactly $N$ qubits (Section~\ref{sec:f2q}), the qubit count remains $N$ throughout.

A complete compilation is defined by four coupled variables:
(i) a \emph{fermion-to-qubit mapping} $\mathcal{M}$ from the PPTT family (Section~\ref{sec:f2q}), producing the Pauli decomposition
$H=\sum_{t=1}^{T}\alpha_t P_t$ with supports $\{S_t\}$;
(ii) a \emph{Pauli ordering} $\pi$, i.e., a permutation of the $T$ terms that induces the support sequence $(S_1,\dots,S_T)$;
(iii) for each term $P_t$, a \emph{parity tree} $\mathcal{T}_t$, namely a rooted
spanning tree over $S_t$ defining the parity CNOT tree of the corresponding Pauli gadget (Section~\ref{sec:pauli_scheduling});
and (iv) a \emph{qubit allocation} $\ell_l:[N]\to\mathcal{C}$ for each slice
$l$ of the synthesized circuit, assigning logical qubits to cores subject to the capacity constraints (Section~\ref{sec:modular}).

Each gadget $e^{-i\theta P_t}$ traverses $\mathcal{T}_t$ twice: a forward parity accumulation followed by a mirrored reverse pass. Therefore, the allocation must co-locate both endpoints of every CNOT edge in both passes. Applying the slice-based cost model~\eqref{eq:transfer_cost} to the resulting $L$-slice circuit, the total inter-core transfer cost for one Trotter step is
\begin{equation}
  \label{eq:objective}
  \mathrm{cost}(\mathcal{M},\pi,\{\mathcal{T}_t\},\{\ell_l\})
  =
  \sum_{l=1}^{L-1}\sum_{q=1}^{N}
  d\!\left(\ell_l(q),\ell_{l+1}(q)\right),
\end{equation}
and the optimization problem is
\begin{equation}
  \label{eq:opt}
  \min_{\mathcal{M},\,\pi,\,\{\mathcal{T}_t\},\,\{\ell_l\}}
  \mathrm{cost}(\mathcal{M},\pi,\{\mathcal{T}_t\},\{\ell_l\}).
\end{equation}

These four variables are tightly coupled. The mapping $\mathcal{M}$ determines the supports $\{S_t\}$ and therefore the candidate parity trees and the qubit-interaction structure seen by the allocator. The ordering $\pi$ controls the smoothness of the support sequence across consecutive gadgets, potentially reducing qubit displacement between slices. Each tree $\mathcal{T}_t$ fixes the CNOT interactions required by the gadget, while the allocation $\{\ell_l\}$ determines the corresponding transfer distances. A suboptimal choice for any one variable constrains the others.

The joint search space grows super-exponentially with $N$. The F2Q search space contains $A^{(3)}_N \cdot N!$ candidates, where
\begin{equation}
  A^{(3)}_N = \frac{1}{2N+1}\binom{3N}{N}
\end{equation}
is the ternary Fuss--Catalan number counting rooted ordered ternary trees with $N$ vertices in the PPTT construction, and $N!$ accounts for mode assignments. Let $T$ denote the number of Pauli terms produced by the mapped Hamiltonian. For electronic-structure Hamiltonians, $T$ is dominated by quartic interaction terms and scales as $O(N^4)$ in the worst case, so the Pauli ordering space has size $T!$. For each Pauli gadget of weight $w$, Cayley's formula gives $w^{\,w-2}$ labelled spanning trees as candidate parity trees. Finally, for each slice, the number of valid qubit-to-core allocations is
\begin{equation}
  \sum_{\substack{x_1+\cdots+x_{|\mathcal{C}|}=N \\ 0 \le x_i \le \kappa}}
  \frac{N!}{x_1!\cdots x_{|\mathcal{C}|}!}.
\end{equation}
The overall solution space is the Cartesian product of these choices, rendering exhaustive search infeasible and motivating the heuristic approach of Section~\ref{sec:methodology}.

\input{methodology}

\section{Evaluation}
\label{sec:experiments}

We evaluate \brand on 14 molecular workloads from the PubChem database~\cite{kim2025pubchem} in the STO-3G basis: water, ammonia, methane, carbon dioxide, urea, glycine, alanine, ethylene, benzene, ethanol, acetone, acetic acid, uracil, and cytosine. The molecular integrals and electronic Hamiltonians are generated with PySCF~\cite{sun2020recent}, after which we convert them into the sparse Majorana representation used by our pipeline, and validate the resulting Hamiltonians against OpenFermion~\cite{mcclean2020openfermion}. These workloads span $14$ to $90$ fermionic modes and approximately $4.2\times 10^3$ to $6.5\times 10^6$ sparse Majorana terms. For each workload, we consider a 2D grid multi-core architecture with fixed core capacity $8$ and choose the smallest rectangular grid with sufficient total capacity, ranging from $(1\!\times\!2,8)$ for the smallest molecules up to $(3\!\times\!4,8)$ for the largest ones.

For the genetic mapping stage we consider population size $50$ and $25$ generations, with transfer cost as the optimization objective. Communication is always reported as the total weighted inter-core transfer cost defined by the target architecture as in \cite{bandic2023mapping,escofet2023hungarian,russo2025optimizing}.

We consider three experiment families. First, we compare our full pipeline against baselines. Second, we isolate the effect of the fermion-to-qubit mapping while holding scheduling and allocation fixed. Third, we isolate the effect of scheduling while holding the mapping and allocation fixed. In all plots, molecules are ordered by increasing number of fermionic modes.

\subsection{Comparison Against Baselines}
\label{sec:exp_baselines}

\begin{figure}
    \centering
    \includegraphics[width=1\linewidth]{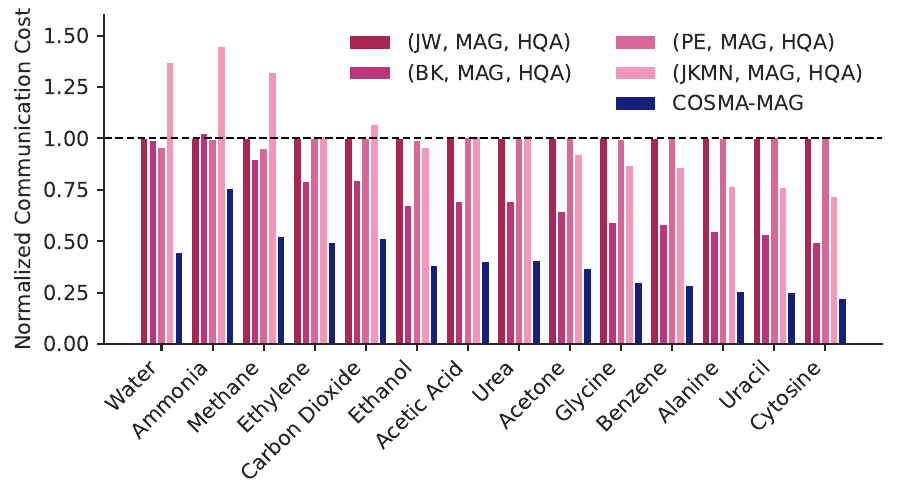}
    \vspace{-0.65cm}
    \caption{Comparison against baselines considering magnitude Pauli scheduling.}
    \label{fig:baseline_mag}
\end{figure}

\begin{figure}
    \centering
    \includegraphics[width=1\linewidth]{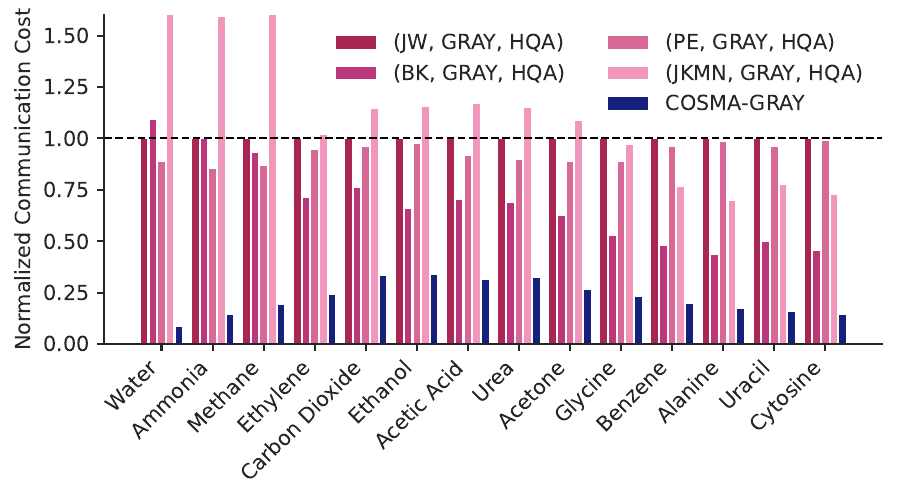}
    \vspace{-0.65cm}
    \caption{Comparison against baselines considering Gray-like Pauli scheduling.}
    \label{fig:baseline_gray}
\end{figure}

The comparison baselines use fixed fermion-to-qubit mappings (JW, BK, PE, and JKMN) combined with two scheduling strategies, namely Gray-inspired and magnitude ordering. For allocation we consider the Hungarian qubit assignment algorithm (HQA)~\cite{escofet2023hungarian}. To make the Hungarian allocator applicable, each Pauli gadget is first synthesized into an index-ordered CNOT chain, where the support qubits are connected from lower index to higher index. This produces an explicit sequence of two-qubit gates, which is then sliced and passed to Hungarian assignment on the same grid architecture. We compare this baseline family against our full COSMA pipeline, which combines the genetic tree-based mapping, same scheduling, and the synthesis and allocation co-optimizer.

As shown in Fig.~\ref{fig:baseline_gray}, relative to the best fixed-mapping Hungarian baseline using Gray-inspired scheduling, the full \brand pipeline attains a median transfer-cost reduction of $59.7\%$. As shown in Fig.~\ref{fig:baseline_mag}, when all methods use magnitude ordering, our tree-based mapping with parity-tree-aware allocation attains a median reduction of $42.8\%$. These improvements show that jointly optimizing mapping, scheduling, and parity-tree-aware allocation is substantially more effective than applying Hungarian assignment after a fixed F2Q mapping and circuit synthesis.

\subsection{Ablation Studies}
\label{sec:exp_ablation}

\begin{figure}
    \centering
    \includegraphics[width=1\linewidth]{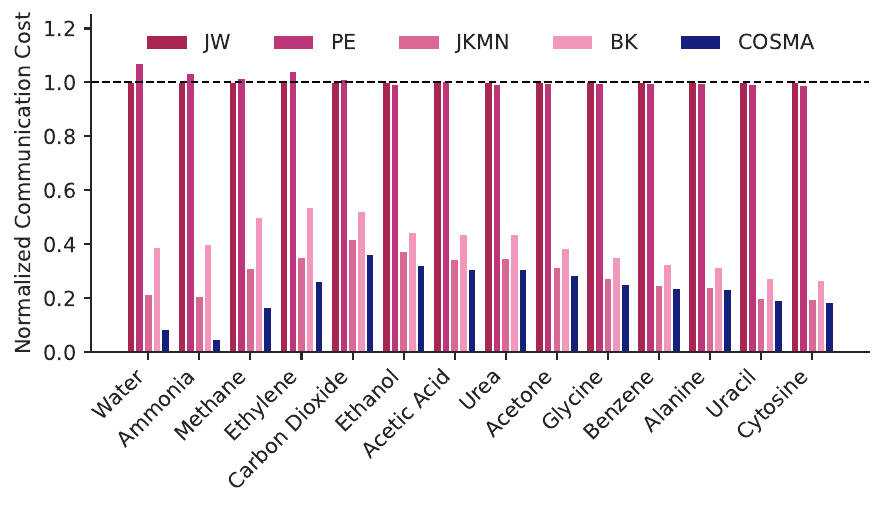}
    \vspace{-0.65cm}
    \caption{Impact of mapping on communication cost.}
    \label{fig:impact_mapping}
\end{figure}

\begin{figure}
    \centering
    \includegraphics[width=1\linewidth]{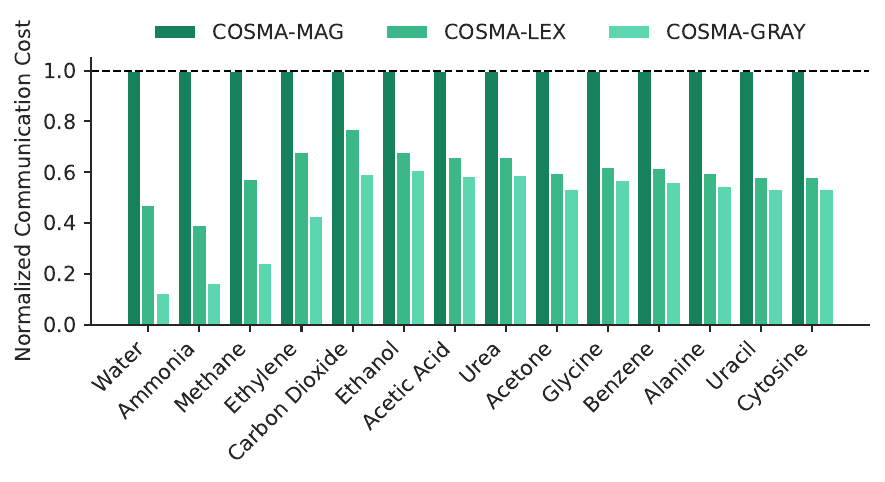}
    \vspace{-0.65cm}
    \caption{Impact of scheduling on communication cost.}
    \label{fig:impact_scheduling}
\end{figure}

\paragraph{Impact of fermion-to-qubit mapping}
To isolate the effect of the mapping alone, we fix the scheduler to Gray-inspired ordering and use the same parity-tree-aware allocation for all mappings. The results of this ablation are shown in Fig.~\ref{fig:impact_mapping}. Relative to the best fixed mapping among JW/BK/PE/JKMN on each workload, the genetic tree-based mapping achieves a median transfer-cost improvement of $11.4\%$. Under this controlled setting, the mapping stage therefore contributes a clear communication benefit. At the same time, the relative reduction tends to become smaller as the number of modes grows: the maximum observed improvement is $76.3\%$ (ammonia, $16$ modes), while the minimum is $3.2\%$ (uracil, $88$ modes). A plausible explanation is that the mapping search space expands rapidly with system size, whereas in our experiments the genetic optimizer uses a fixed budget of $25$ generations and population size $50$ for all molecules. As a result, the optimizer likely explores a smaller fraction of the candidate mappings for the larger instances, which can reduce the gains attainable from mapping optimization alone.

\paragraph{Impact of scheduling}
To isolate scheduling, we fix the genetic mapping and the parity-tree-aware allocator, and compare Gray-inspired, lexicographic, and magnitude ordering. As shown in Fig.~\ref{fig:impact_scheduling}, Gray-inspired scheduling is consistently the strongest of the three: relative to lexicographic ordering, the median transfer-cost ratio is $0.89$ (median reduction $10.5\%$), while relative to magnitude ordering the median ratio is $0.54$ (median reduction $46.0\%$). This confirms that explicitly smoothing support transitions is strongly aligned with the communication objective, whereas coefficient-magnitude ordering is poorly suited to multi-core execution.

\subsection{Runtime}
\label{sec:exp_runtime}

\begin{figure}
    \centering
    \includegraphics[width=1\linewidth]{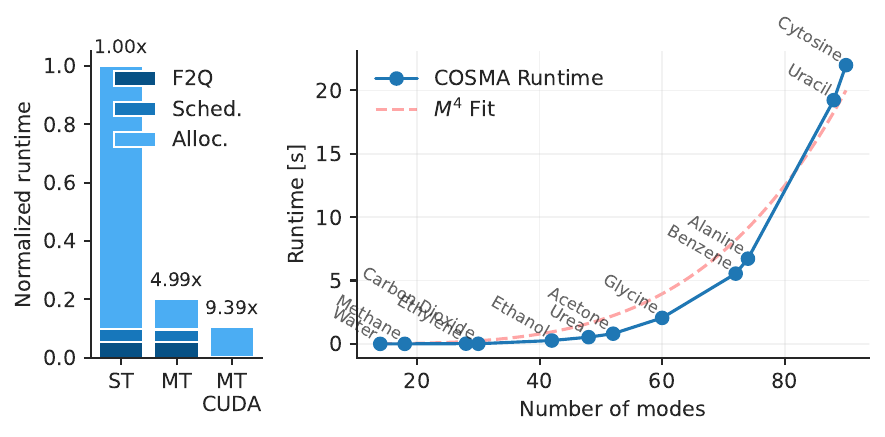}
    \caption{Single-thread (ST), multi-thread (MT) and CUDA accelerated implementation runtime comparison for Ethylene (left). COSMA runtime for different numbers of modes (right).}
    \label{fig:runtime}
\end{figure}

COSMA is implemented in C++, and experiments were run on a machine equipped with an AMD Ryzen 5900X \mbox{12-cores} CPU and an RTX 6000 Ada GPU. Since the proposed approach uses a genetic algorithm, fast inner-loop evaluation is essential to explore many candidate solutions in reasonable time. For this reason, we investigated CUDA acceleration~\cite{cuda} for parts of the pipeline, including F2Q mapping (Pauli-string replacement and accumulation) and comparison-based sorting (magnitude, lexicographic, and Gray-inspired) via Thrust~\cite{bell2012thrust}.

The synthesis-and-allocation stage is intrinsically sequential and dominates total runtime. To reduce wall-clock time, we adopt a divide-and-conquer approach: the Pauli sequence is partitioned into chunks (one per CPU thread), each chunk is allocated in parallel from an initial layout, and chunks are then stitched together while accounting for transition communication between boundary layouts. This preserves optimization quality while improving throughput.

Fig.~\ref{fig:runtime} (left) reports runtime on ethylene for the JW mapping, Gray-inspired scheduling, and COSMA allocation pipeline. Multi-threaded CPU execution provides a $5.0\times$ speedup over single-threaded CPU, whereas CUDA provides a $9.4\times$ speedup over single-threaded CPU and a further $1.88\times$ speedup over the 24-thread CPU run.

More broadly, Fig.~\ref{fig:runtime} (right) shows that in the CUDA implementation runtime grows from approximately $2$\,ms (water) to approximately $22$\,s (cytosine). The scalability trend is well approximated by a quartic fit in the number of modes: allocation time scales approximately linearly with the number of Pauli terms, while the number of Pauli terms itself scales approximately quartically with the number of modes. Consequently, there is no evidence of additional super-linear scaling beyond Pauli-term growth.

\section{Limitations and Future Work}
\label{sec:limitations}

Several aspects of the present work invite further investigation. Similarly to other works, we adopt a generic transfer cost model based on shortest-path inter-core distance~\cite{escofet2023hungarian,bandic2023mapping}, which applies to both multi-core and distributed quantum systems but does not account for link capacities, congestion, or latency asymmetries. A more detailed model reflecting specific hardware platforms is left for future work~\cite{cacciapuoti2026quantum,caleffi2024distributed}.

Our synthesis treats consecutive Pauli gadgets independently. Exploiting cancellations between the reverse CNOT tree of one gadget and the forward tree of the next could reduce the number of inter-core interactions~\cite{li2022paulihedral}, and incorporating commuting-group structure~\cite{gui2020term} into the scheduling stage could further improve both gate count and Trotter error. More broadly, we currently ignore the Trotter error introduced by reordering non-commuting terms; quantifying and jointly optimizing the trade-off between simulation accuracy and communication cost is an important direction for future work.

The genetic algorithm requires evaluating the full scheduling and allocation pipeline for each candidate mapping, which dominates runtime. Faster surrogate objectives based on interaction-weighted Pauli weight or support-overlap statistics could accelerate the search, and more expressive evolutionary operators remain to be explored. This work also restricts attention to PPTT mappings, which encode $N$ fermionic modes onto exactly $N$ qubits. Mappings outside this family, such as those based on quantum low-density parity-check codes, allow different mode-to-qubit scaling ratios and may yield sparser interaction graphs more favourable for modular architectures~\cite{maskara2025fast,gandon2025stabilizer,chiew2025optimal}.

Finally, the current objective minimises inter-core transfer count without considering whether communication can be overlapped with local computation. Pipelining the two could reduce wall-clock simulation time and is a natural extension for distributed architectures. Extending the evaluation to lattice models such as the Fermi-Hubbard and Sachdev-Ye-Kitaev models, as well as variational algorithms beyond Trotterised simulation and modular architectures beyond meshes, would further broaden the framework's applicability and stress-test the heuristics on qualitatively different interaction structures.

\section{Related Work}
\label{sec:related}

Prior work relevant to this paper spans three main areas: fermion-to-qubit mappings, Pauli-level compilation of simulation kernels, and compilation for modular or multi-core quantum architectures.

\paragraph{Fermion-to-qubit mappings}
Standard encodings such as Jordan--Wigner and Bravyi--Kitaev remain widely used baselines for quantum simulation~\cite{whitfield2011simulation,tranter2018comparison,jiang2020optimal}. More recent works optimize the mapping itself, including ternary-tree constructions~\cite{jiang2020optimal} and adaptive approaches such as Bonsai, HATT, TOPP-HATT and Treespilation~\cite{miller2023bonsai,liu2025hatt,de2025optimised,miller2026treespilation}. Other methods formulate mapping as a global optimization problem~\cite{liu2024fermihedral,chiew2025optimal}. While these works improve Pauli weight, depth, or locality, they do not target inter-core communication in modular architectures.

\paragraph{Pauli scheduling and compilation}
The ordering and synthesis of Pauli operators significantly affect both circuit cost and simulation accuracy~\cite{tranter2019ordering,childs2019theory}. Approaches based on commutativity and TSP-like formulations improve execution ordering and error mitigation~\cite{gui2020term,tomesh2021optimized}, while frameworks such as Paulihedral, PauliForest, and Tetris exploit Pauli structure to reduce gate count and routing overhead~\cite{li2022paulihedral,li2024pauliforest,jin2024tetris}. Recent work also revisits lexicographic strategies for improved Pauli ordering under compilation constraints~\cite{huang2024redefining}. Relatedly, partial Trotterization has recently been proposed as a compiler optimization to reduce Hamiltonian-simulation cost by selectively applying product-form decompositions~\cite{decker2025kernpiler}. However, these methods primarily optimize gate-level metrics and local connectivity rather than communication across multiple quantum cores.

\paragraph{Modular and distributed quantum compilation}
Modular architectures have been proposed to address scalability limitations of monolithic devices~\cite{monroe2014large,jnane2022multicore,rodrigo2021double}. This has motivated work on circuit partitioning, qubit allocation, and routing for multi-core systems, including time-sliced partitioning~\cite{baker2020time}, Hungarian-based assignment~\cite{escofet2023hungarian}, and QUBO or heuristic mapping approaches~\cite{bandic2023mapping,escofet2024revisiting,escofet2024route,kaur2025optimized}. At a higher level, distributed quantum simulation and execution models further highlight communication as a primary bottleneck~\cite{feng2024distributed,buessen2023simulating}. These works typically optimize placement and routing for a fixed circuit.

\section{Conclusion}
\label{sec:conclusion}

We presented \brand, a communication-aware compilation framework for fermionic simulation on modular quantum architectures. By jointly optimizing fermion-to-qubit mapping, Pauli scheduling, and qubit allocation, our approach significantly reduces inter-core communication compared to conventional pipelines. Unlike prior work, which typically optimizes these stages in isolation or focuses on single-core metrics, \brand explicitly targets communication through a cross-layer co-design. These results highlight the importance of integrating mapping, scheduling, and allocation to achieve scalable quantum simulation on multi-core systems.

\section*{Acknowledgements}

The authors gratefully acknowledge funding from the European Commission through HORIZON-EIC-2022-PATHFINDEROPEN-01-101099697 (QUADRATURE). The authors also gratefully acknowledge NVIDIA Corporation for the donation of GPU hardware used in this work.

\bibliographystyle{bib/IEEETran}
\bibliography{bib/IEEEabrv,references}

\end{document}

%% file: background.tex
\section{Background}
\label{sec:background}

Fig.~\ref{fig:background} summarises the steps required to evaluate the communication cost of quantum simulation starting from a target molecule and a modular quantum architecture. This section reviews the main concepts used throughout the paper for each of the steps of Fig.~\ref{fig:background}. 

\subsection{Quantum Simulation}
\label{sec:quantum_simulation}

Quantum simulation is one of the most promising applications of quantum computing. The central idea is to use a controllable quantum system to reproduce the dynamics of another quantum system that is intractable on classical hardware. Many systems of interest, such as molecules and strongly correlated materials, are described by Hamiltonians acting on Hilbert spaces whose dimension grows exponentially with the number of particles. In electronic-structure problems, this leads to computational costs that grow combinatorially with system size and quickly become prohibitive for classical methods~\cite{gui2020term,szabo2012modern}.

Simulating a fermionic system generally involves three steps: state preparation, time evolution, and measurement of observables~\cite{gui2020term,whitfield2011simulation}. This work focuses on compiling the time-evolution operator for modular quantum computers.

The dynamics of a closed quantum system are governed by the time-dependent Schrödinger equation
\begin{equation}
  i\hbar\,\frac{\partial}{\partial t}\ket{\psi(t)} = H\ket{\psi(t)},
\end{equation}
where $H$ is the system Hamiltonian. Its formal solution is
\begin{equation}
  \ket{\psi(t)} = e^{-iHt}\ket{\psi(0)},
\end{equation}
where $U(t)=e^{-iHt}$ is the time-evolution operator. On a gate-based quantum computer, this operator must be approximated by a sequence of elementary native gates.

In digital quantum simulation, the Hamiltonian is decomposed into a sum of Pauli strings,
\begin{equation}
  \label{eq:pauli_expansion}
  H = \sum_j \alpha_j P_j,
\end{equation}
where $P_j$ are Pauli strings and $\alpha_j \in \mathbb{R}$. A \emph{Pauli string} acting on $n$ qubits is a tensor product of single-qubit Pauli operators,
\begin{equation}
  P_j = \bigotimes_{k=1}^{n} \sigma_{j,k}, \qquad \sigma_{j,k} \in \{I, X, Y, Z\}.
\end{equation}
The \emph{support} of $P_j$ is the set of qubits on which it acts non-trivially,
\[
  S_j = \mathrm{supp}(P_j) := \bigl\{\, k \mid \sigma_{j,k} \neq I \,\bigr\},
\]
and its \emph{Pauli weight} is $\mathrm{weight}(P_j):=|S_j|$.

The time-evolution operator can be approximated using the first-order Trotter--Suzuki product formula
\begin{equation}
  \label{eq:trotter}
  e^{-iHt} \approx \biggl(\prod_j e^{-i\alpha_j P_j\,t/r}\biggr)^{r},
\end{equation}
where $r$ is the number of Trotter steps. Each factor $e^{-i\alpha_j P_j\,t/r}$, called a \emph{Pauli gadget}, can be implemented using a structured gate sequence described in Section~\ref{sec:pauli_scheduling}.

Beyond product formulas, methods such as qubitization and quantum signal processing offer improved asymptotic scaling~\cite{low2019hamiltonian}.

A central application of quantum simulation is quantum chemistry, where the goal is to predict molecular properties by solving the electronic Schrödinger equation. Under the Born--Oppenheimer approximation, nuclear coordinates are treated as fixed parameters and the electronic Hamiltonian takes the form (in atomic units)
\[
  H_{\mathrm{elec}}
  = -\sum_{j\in[M]}\frac{\nabla_{\mathbf{r}_j}^{2}}{2}
    -\sum_{\substack{i\in[N]\\j\in[M]}}\frac{Z_i}{|\mathbf{R}_i-\mathbf{r}_j|}
    +\sum_{\substack{j,k\in[M]\\j<k}}\frac{1}{|\mathbf{r}_j-\mathbf{r}_k|},
\]
where $\mathbf{R}_i$ and $Z_i$ are the position and charge of the $i$-th nucleus, $\mathbf{r}_j$ is the position of the $j$-th electron, $N$ is the number of nuclei, and $M$ is the number of electrons.

To make this problem amenable to quantum algorithms, the Hamiltonian is expressed in second quantisation. The electronic wavefunction is expanded in a finite basis of $K$ spin-orbitals $\{\chi_p\}_{p\in[K]}$, where $\chi_p(x)$ depends on combined spatial and spin coordinates $x=(\mathbf{r},\sigma)$. Introducing fermionic creation and annihilation operators $a_p^\dagger$ and $a_p$ satisfying the canonical anticommutation relations
\begin{equation}
  \label{eq:anticommutation}
  \{a_p,\, a_q^\dagger\} = \delta_{pq}\,\mathds{1}, \qquad
  \{a_p,\, a_q\} = \{a_p^\dagger,\, a_q^\dagger\} = 0,
\end{equation}
the electronic-structure Hamiltonian becomes
\begin{equation}
  \label{eq:mol_hamiltonian}
  H = \sum_{p,q\in[K]} h_{pq}\,a_p^\dagger a_q
    + \frac{1}{2}\sum_{p,q,r,s\in[K]} g_{pqrs}\,a_p^\dagger a_q^\dagger a_s a_r,
\end{equation}
where $h_{pq}$ are one-electron integrals encoding kinetic energy and electron--nuclear attraction, and $g_{pqrs}$ are two-electron Coulomb integrals encoding electron--electron repulsion. This Hamiltonian is the starting point for quantum algorithms such as the variational quantum eigensolver (VQE) and quantum phase estimation.

Since quantum computers operate on qubits rather than fermionic modes, a mapping between these two representations is required.

\subsection{Fermion-to-Qubit Mapping}
\label{sec:f2q}

Fermion-to-qubit (F2Q) mappings provide a systematic way to encode fermionic operators as qubit operators. This step is necessary because fermionic operators satisfy the anticommutation relations~\eqref{eq:anticommutation}, whereas qubit operators obey the Pauli algebra. The goal is to represent each pair $(a_p^\dagger, a_p)$ as Pauli strings while preserving the fermionic algebra.

An $N$-mode fermionic system is described by operators $\{a_p, a_p^\dagger\}_{p\in[N]}$ satisfying \eqref{eq:anticommutation}. The corresponding state space is the Fock space $\mathcal{F}(\mathbb{C}^N)$, a $2^N$-dimensional Hilbert space spanned by the Fock basis. The \emph{fermionic vacuum} $\ket{\Omega}$ is the unique state annihilated by all annihilation operators,
\[
  a_p\ket{\Omega} = 0 \qquad \forall\, p\in[N].
\]
All other basis states are obtained by applying creation operators,
\[
  \ket{n_1 n_2 \cdots n_N} := \prod_{p=1}^{N}(a_p^\dagger)^{n_p}\ket{\Omega},
\]
for occupation numbers $n_p\in\{0,1\}$.

A convenient intermediate representation uses the $2N$ \emph{Majorana operators} $\{m_k\}_{k\in[2N]}$, defined by
\begin{equation}
  \label{eq:majorana_def}
  a_p^\dagger = \frac{m_{2p-1} - i\,m_{2p}}{2}, \qquad
  a_p         = \frac{m_{2p-1} + i\,m_{2p}}{2},
\end{equation}
for all $p\in[N]$. These operators are Hermitian and satisfy
\[
  \{m_i,m_j\}=2\delta_{ij}\,\mathds{1},
\]
which implies $m_i^2=\mathds{1}$. In terms of Majorana operators, the electronic Hamiltonian~\eqref{eq:mol_hamiltonian} can be rewritten as
\begin{equation}
  \label{eq:hamiltonian_majorana}
  H = \sum_{p,q\in[2N]} i\,c_{pq}\,m_p m_q
    + \sum_{p,q,r,s\in[2N]}\!\! c_{pqrs}\,m_p m_q m_r m_s,
\end{equation}
for suitable coefficients $c_{pq}$ and $c_{pqrs}$.

The Fock space $\mathcal{F}(\mathbb{C}^N)$ and the $N$-qubit Hilbert space $\bigotimes_{p=1}^{N}\mathbb{C}^2$ are both isomorphic to $\mathbb{C}^{2^N}$, so a unitary encoding between them always exists. The simplest choice encodes occupation numbers directly as qubit states by identifying each Fock basis vector $\ket{n_1\cdots n_N}$ with the corresponding computational basis state. This encoding, the \emph{Jordan--Wigner (JW) transformation}, maps the Majorana operators for all $p\in[N]$ to Pauli strings as
\begin{equation}
  \label{eq:jw}
  m_{2p-1} \;\mapsto\; X_p \prod_{k=1}^{p-1} Z_k, \qquad
  m_{2p}   \;\mapsto\; Y_p \prod_{k=1}^{p-1} Z_k,
\end{equation}
where $\sigma_k$, for $\sigma\in\{X,Y,Z\}$, denotes $\sigma$ acting on qubit $k$ and identity on all others.

More generally, a \emph{Pauli-string F2Q mapping} assigns a Pauli string $P_k$ to each Majorana operator $m_k$ such that the anticommutation algebra is preserved:
\begin{equation}
  \label{eq:majorana_strings}
  \{m_i, m_j\} = 2\delta_{ij}\,\mathds{1}
  \quad\mapsto\quad
  \{P_i, P_j\} = 2\delta_{ij}\,\mathds{1}.
\end{equation}
Jordan--Wigner belongs to the broader class of \emph{product-preserving ternary-tree (PPTT)} F2Q mappings~\cite{miller2023bonsai}. Mappings in this class share several useful properties: they map Fock-basis product states to computational-basis states, send the fermionic vacuum to $\ket{0}^{\otimes N}$, and map the Hartree--Fock state to a computational-basis state. These are important reference states in many quantum chemistry algorithms, and their preparation therefore requires no entangling gates.

\begin{figure}
  \centering
  \includegraphics{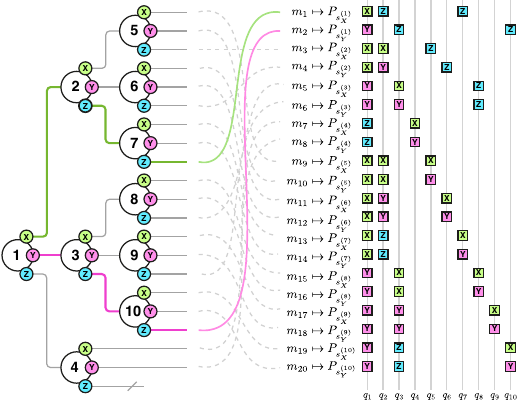}
  \caption{JKMN-like~\cite{jiang2020optimal} ternary-tree mapping example for 10 modes with identity mode-order bijection $f(u)=u$.}
  \label{fig:ternary_tree}
\end{figure}

The \emph{Bonsai} framework~\cite{miller2023bonsai} provides a unified combinatorial construction for PPTT mappings. The mapping is represented by a rooted ternary tree whose $N$ vertices are in bijection with the $N$ qubits, as shown in Fig.~\ref{fig:ternary_tree}. Each vertex $u$ has up to three outgoing downward links labelled $X$, $Y$, and $Z$; missing links are completed with \emph{legs} (dangling edges), so that every vertex has exactly three outgoing links. A counting argument shows that an $N$-node tree completed in this way has $2N+1$ legs.

Each leg defines a unique rootward path. A Pauli string is obtained by following this path and recording the Pauli label of each crossed link; the resulting operator acts non-trivially only on the qubits visited by the path. Any two rootward paths share a first common ancestor at which the corresponding strings carry different non-identity Pauli labels, so the corresponding Pauli strings anticommute. After discarding one redundant string, the remaining $2N$ strings can be identified with the $2N$ Majorana operators, yielding a valid Majorana-string F2Q mapping.

To make the mapping product-preserving, Bonsai specifies a pairing rule: for each vertex $u$, follow its $X$-labelled link and then recursively follow $Z$-labelled links until a leg $s_X^{(u)}$ is reached; the same procedure starting from the $Y$-labelled link yields $s_Y^{(u)}$. The Majorana operators of one fermionic mode are then assigned to the strings associated with these two legs:
\begin{equation}
  m_{2p-1} \mapsto P_{s_X^{(u)}}, \qquad m_{2p} \mapsto P_{s_Y^{(u)}},
\end{equation}
where $p = f(u)$ for a mode-order bijection $f$ between tree vertices and fermionic modes, as shown in Fig.~\ref{fig:ternary_tree} for \mbox{$p=u=1$}. This rule guarantees that the fermionic vacuum maps to $\ket{0}^{\otimes N}$ and that important reference states can be prepared without entanglement~\cite{miller2023bonsai}. 

The PPTT formalism recovers standard encodings as special cases. Jordan--Wigner~(JW)~\cite{jordan1928paulische} is obtained from a \mbox{$Z$-chain} tree, parity encoding~(PE)~\cite{bravyi2017tapering} from a \mbox{$X$-chain}, JKMN~\cite{jiang2020optimal} corresponds to a complete ternary tree, and Braviy--Kitaev~(BK)~\cite{bravyi2002fermionic} corresponds to a specific tree instance.

\emph{Treespilation}~\cite{miller2026treespilation} treats the PPTT mapping itself as an optimisation variable. Rather than fixing an encoding a priori, Treespilation searches over the space of PPTT mappings to minimise a user-defined cost function, such as total Pauli weight, entangling-gate count, or transpilation cost on a target hardware topology. Starting from an initial PPTT mapping, local structure-preserving transformations are applied, including moving leaves to free legs, changing the root, and permuting fermionic modes. Simulated annealing is then used to explore the resulting discrete search space. In this way, Treespilation adapts the F2Q mapping to both the operator set and the hardware constraints.

\begin{figure}
  \centering
  \includegraphics[width=\mycolwidth]{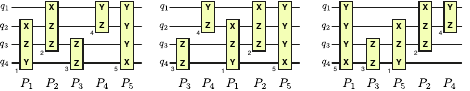}
  \caption{Three possible orderings of Pauli strings, showing their supports for a simple one-dimensional qubit ordering.}
  \label{fig:scheduling}
\end{figure}

\begin{figure}
  \newcommand{\qtkzga}[1]{%
    \gategroup[wires=5,steps=2,style={dashed,rounded corners,fill=green!10,inner xsep=2pt},background,
      label style={label position=above,anchor=north,yshift=0.7cm}]{\textsc{#1}}}
  \newcommand{\qtkzgb}[1]{%
    \gategroup[wires=5,steps=2,style={dashed,rounded corners,fill=cyan!10,inner xsep=2pt},background,
      label style={label position=below,anchor=north,yshift=-0.3cm}]{\textsc{#1}}}
  \newcommand{\qtkzgc}[1]{%
    \gategroup[wires=5,steps=1,style={dashed,rounded corners,fill=magenta!10,inner xsep=2pt},background,
      label style={label position=above,anchor=north,yshift=0.7cm}]{\textsc{#1}}}
  \centering
  \resizebox{\mycolwidth}{!}{
  \begin{quantikz}[row sep=0.15cm]
    \lstick{$q_1$} & \gate[5]{e^{-i\theta P}} & \rstick{} \\
    \lstick{$q_2$} & \phantomgate{Y}          & \rstick{} \\
    \lstick{$q_3$} & \phantomgate{Y}          & \rstick{} \\
    \lstick{$q_4$} & \phantomgate{Y}          & \rstick{} \\
    \lstick{$q_5$} & \phantomgate{Y}          & \rstick{}
  \end{quantikz}
  =
  \begin{quantikz}[row sep=0.15cm]
    \lstick{} & \phantomgate{Y}\qtkzga{\shortstack{1-qubit\\gates}} &[-0.3cm]
              & \ctrl{1}\qtkzgb{\shortstack{CNOT tree\\\small{\textit{parity circuit}}}}
              &
              & \qtkzgc{\shortstack{root\\rotation}}
              & \qtkzgb{\shortstack{mirrored\\CNOT tree}}
              & \ctrl{1} & \qtkzga{\shortstack{1-qubit\\gates}} &[-0.3cm] & \rstick{} \\
    \lstick{} & \gate{S^\dagger} & \gate{H}        & \targ{}   & \ctrl{1} & & \ctrl{1} & \targ{}   & \gate{H}       & \gate{S} & \rstick{} \\
    \lstick{} & \phantomgate{Y}  &                 & \targ{}   & \targ{}  & \gate{R_Z(2\theta)} & \targ{} & \targ{} & \phantomgate{Y} & & \rstick{} \\
    \lstick{} & \gate{H}         &                 & \ctrl{-1} &          & &          & \ctrl{-1} &                 & \gate{H} & \rstick{} \\
    \lstick{} & \phantomgate{Y}  &                 &           &          & &          &           & \phantomgate{Y} &          & \rstick{}
  \end{quantikz}
  }
  \caption{Gate-level implementation of the Pauli gadget $e^{-i\theta P}$ for $P = Z_1 Y_2 Z_3 X_4 I_5$.}
  \label{fig:kernel}
\end{figure}

After applying a Pauli-string F2Q mapping, each Majorana operator $m_k$ is replaced by its corresponding Pauli string $P_k$. Since any product of Pauli strings equals, up to a global phase, another Pauli string, every fermionic monomial in \eqref{eq:hamiltonian_majorana} maps to a Pauli-string operator on qubits. The fermionic Hamiltonian therefore takes the form of \eqref{eq:pauli_expansion}, with coefficients $\alpha_j$ absorbing both the original fermionic coefficients and the phases arising from Pauli multiplication.

Different F2Q mappings produce different Pauli decompositions of the same fermionic Hamiltonian, altering the support and Pauli weight of the resulting strings and therefore the cost of implementing the corresponding operations. On modular or limited-connectivity hardware, the spatial distribution of the support is equally important, since strings spanning distant qubits or multiple cores incur additional routing overhead. This motivates architecture-aware optimisation of the F2Q mapping.

\subsection{Pauli Scheduling}
\label{sec:pauli_scheduling}

Given the Pauli decomposition~\eqref{eq:pauli_expansion} and the Trotter approximation~\eqref{eq:trotter}, implementing a single Trotter step reduces to implementing a sequence of Pauli gadgets
\[
e^{-i\alpha_j P_j t/r} = e^{-i\theta_j P_j},
\qquad \theta_j := \alpha_j t/r.
\]
As illustrated in Fig.~\ref{fig:kernel}, each gadget $e^{-i\theta P}$ can be realised in five stages:
\begin{enumerate*}[(i)]
  \item single-qubit basis changes diagonalise each non-identity factor of $P$ into the $Z$ basis: apply $H$ where $\sigma_k=X$, and $S^\dagger H$ where $\sigma_k=Y$;
  \item a CNOT tree accumulates the parity of all qubits in $\mathrm{supp}(P)$ into a designated \emph{root} qubit;
  \item an $R_Z(2\theta)$ gate is applied to the root qubit;
  \item the CNOT tree is applied in reverse order to uncompute the parity accumulation;
  \item the single-qubit basis changes are inverted.
\end{enumerate*}

\begin{figure}[t!]
  \centering
  \vspace{-0.5cm}
  \includegraphics[width=\linewidth]{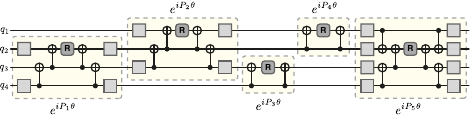}
  \caption{Synthesized circuit for one Trotter step after decomposing each Pauli gadget as in Fig.~\ref{fig:kernel}.}
  \label{fig:synthesis}
\end{figure}

For large Hamiltonians, the number of Pauli terms can be very large, and the ordering of gadgets in the Trotter product significantly affects circuit cost. Two Pauli gadgets $e^{-i\theta_a P_a}$ and $e^{-i\theta_b P_b}$ commute, and can therefore be exchanged freely, if and only if $[P_a,P_b]=0$.

Although reordering non-commuting terms does not change the formal order of the Trotter error, it can still affect the constant factors and practical error magnitude~\cite{grimsley2019adaptive,huang2024redefining}. Beyond the error perspective, ordering also strongly influences the amount of gate cancellation achievable during synthesis. Several strategies have therefore been proposed, including grouping commuting terms into blocks simulated exactly within each group~\cite{gui2020term}, lexicographic ordering as a heuristic for maximising cancellation opportunities~\cite{tranter2018comparison,hastings2014improving}, and joint optimisation of gate count and Trotter-error bounds~\cite{cowtan2020generic,gui2020term}.

In this work, following~\cite{li2022paulihedral,huang2024redefining}, we do not explicitly optimise reordering-induced Trotter error and instead focus on minimising inter-core communication. This choice is motivated by our target setting, where communication overhead is expected to dominate overall execution cost, while Trotter error can be reduced independently through smaller time steps or higher-order product formulas~\cite{low2019well}.

Fig.~\ref{fig:scheduling} shows three possible orderings of Pauli terms in the qubit Hamiltonian, highlighting the qubits in their supports. Fig.~\ref{fig:synthesis} shows the resulting synthesized circuit for the first one of such orderings after each Pauli gadget has been implemented as in Fig.~\ref{fig:kernel}. Crucially, the Pauli ordering together with the choice of parity trees determines the resulting sequence of two-qubit interactions.

\subsection{Modular Quantum Computers}
\label{sec:modular}

\begin{figure}
     \centering
     \begin{subfigure}[b]{0.47\linewidth}
         \centering
         \includegraphics{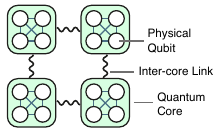}
         \caption{Modular hardware topology.}
         \label{fig:hardware}
     \end{subfigure}
     \hfill
     \begin{subfigure}[b]{0.47\linewidth}
         \centering
         \includegraphics{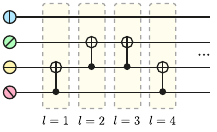}
         \caption{Synthesized circuit slices.}
         \label{fig:slices}
     \end{subfigure}

     \vspace{0.2cm}

     \begin{subfigure}[b]{\linewidth}
         \centering
         \includegraphics[width=\linewidth]{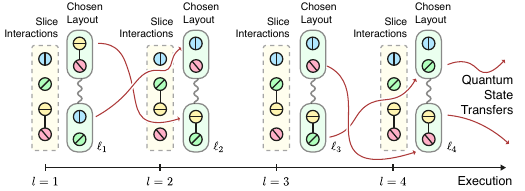}
         \caption{Per-slice allocation decision and execution.}
         \label{fig:allocation}
     \end{subfigure}

    \caption{Modular quantum-architecture model and allocation process for a sliced circuit.}
    \label{fig:modular}
\end{figure}

Scaling quantum processors to millions of qubits is unlikely to be achieved with monolithic architectures alone. Modular quantum computing instead proposes interconnecting multiple QPUs, or cores, through classical and quantum communication links. As shown in Fig.~\ref{fig:hardware}, each core contains a set of tightly coupled qubits supporting high-fidelity local operations, while inter-core communication is typically slower and less reliable.

Qubit states are transferred between modules via communication links supporting both classical and quantum signalling~\cite{rodrigo2021modelling}. Transfer can be realised through quantum teleportation, which consumes a shared Bell pair, or through direct remote-gate mechanisms~\cite{cuomo2023optimized}. Although some hardware platforms may support remote gates natively, in this work we assume that a two-qubit gate can be executed only when both logical qubits reside in the same core~\cite{bandic2023mapping}.

A common first step in qubit-allocation optimisation, for both single-core~\cite{nannicini2022optimal} and multi-core architectures~\cite{baker2020time,bandic2023mapping,escofet2024revisiting}, is to partition the circuit into \emph{slices} of gates that share no logical qubits and can therefore be executed in parallel. Fig.~\ref{fig:slices} shows the first four slices of the synthesized circuit in Fig.~\ref{fig:synthesis}. For each slice, logical qubits are assigned to cores subject to two constraints:
\begin{enumerate*}[(a)]
  \item every pair of qubits involved in the same gate is co-located in the same core, and
  \item the number of logical qubits assigned to each core does not exceed its physical capacity.
\end{enumerate*}

The objective is to minimise the total inter-core communication cost incurred when qubits migrate between cores across consecutive slices. Let $c_{q,l}$ denote the core assigned to logical qubit $q$ in slice $l$, let $Q$ be the total number of logical qubits, $L$ the total number of slices, and let $\mathbf{D}$ be the inter-core distance matrix. According to this model~\cite{bandic2023mapping,escofet2023hungarian}, the total transfer cost is
\begin{equation}
  \label{eq:transfer_cost}
  \sum_{l=1}^{L-1}\sum_{q=1}^{Q} \mathbf{D}[c_{q,l},\, c_{q,l+1}].
\end{equation}

An assignment of logical qubits to cores is also referred to as a \emph{layout}, and we denote the layout of slice $l$ by $\ell_l$. The sequence of layouts across slices defines the qubit-routing process, as it determines how logical qubits move across cores during circuit execution.

Fig.~\ref{fig:allocation} shows a simple case in which the four slices of Fig.~\ref{fig:slices} are allocated on a two-core architecture. During execution, changes in layout correspond to inter-core qubit transfers.

%% file: methodology.tex
\section{Methodology}
\label{sec:methodology}

\begin{figure}
    \centering
    \includegraphics[width=1\linewidth]{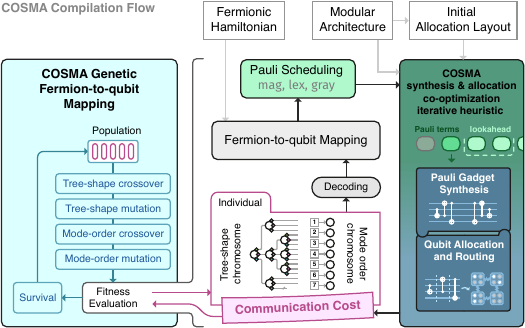}
    \caption{Overview of the COSMA framework.}
    \label{fig:cosma}
\end{figure}

We propose \brand, a framework for Hamiltonian simulation workloads targeting multi-core quantum architectures. The framework jointly optimises three coupled stages: (i)~fermion-to-qubit mapping, (ii)~Pauli-term scheduling, and (iii)~parity-tree synthesis and multi-core qubit allocation. An overview of the pipeline is shown in Fig.~\ref{fig:cosma}. The overall objective is to reduce inter-core state transfers while preserving the while preserving the Hamiltonian semantics.

Starting from a Majorana Hamiltonian~\eqref{eq:hamiltonian_majorana}, \brand selects a fermion-to-qubit transformation, expands and aggregates the resulting Pauli terms, imposes an execution order on them, and finally performs a topology-aware allocation that simultaneously synthesises the CNOT parity tree for each Pauli gadget and manages qubit placement across cores. Let $(P_1,\dots,P_T)$ denote the ordered sequence of Pauli terms, $S_t = \mathrm{supp}(P_t)$ its support, and $M_t \in \{0,1\}^n$ its support mask. The three stages are coupled: the mapping determines the possible supports; the scheduling determines the temporal order in which the sets $S_t$ are encountered; and the allocation stage both synthesises the parity circuit for each term and determines the inter-core transfers needed to make every CNOT executable.

\subsection{Fermion-to-Qubit Mapping}
\label{sec:method_f2q}

We consider both standard F2Q mappings and optimised tree-based mappings as baselines. The standard baselines are JW, BK, PE, and JKMN mappings. Beyond these baselines, we optimise the ternary tree shape and the assignment of fermionic modes to tree nodes using a genetic algorithm~\cite{mitchell1998introduction}.

\paragraph{Genome encoding}
A candidate mapping is encoded as a two-component genome:
\begin{enumerate*}[(a)]
  \item a \emph{tree-shape chromosome}, represented as the preorder degree sequence of an ordered ternary tree, and
  \item a \emph{mode-assignment chromosome}, represented as a permutation of the $N$ fermionic modes over the $N$ tree nodes.
\end{enumerate*}
This factorisation separates structural optimisation from mode placement.

\paragraph{Initialization}
The initial population is seeded with the canonical constructions JW, PE, and JKMN and randomly generated candidates.

\paragraph{Genetic operators}
At each generation, parents are selected by tournament selection. For the tree-shape chromosome, a subtree-swap crossover exchanges equally sized subtrees between two parents, and a subtree-replacement mutation replaces a subtree with a random subtree of equal size. For the mode-assignment chromosome, ordered crossover~\cite{davis1985applying} ensures offspring remain valid permutations, and a swap mutation facilitates exploration. Elitism preserves the best individuals unchanged across generations.

\paragraph{Fitness}
The fitness of a candidate mapping is the total inter-core transfer cost produced by running the downstream scheduling and allocation stages (Sections~\ref{sec:method_scheduling} and~\ref{sec:method_allocation}) with that mapping fixed. The best individual at termination defines the final mapping used downstream.

\subsection{Pauli Scheduling}
\label{sec:method_scheduling}

Once a mapping has been fixed, each Majorana monomial is expanded into a Pauli term via the Majorana strings~\eqref{eq:majorana_strings}, and equal Pauli terms are aggregated by coefficient summation. Terms with negligible coefficient magnitude are discarded, yielding a sparse qubit Hamiltonian with support sets $\{S_j\}_{j}$.

We evaluate three ordering policies for the resulting Pauli terms before allocation.

\paragraph{Magnitude ordering.} Terms are ordered by decreasing coefficient magnitude, a strategy that has been shown to provide low Trotterization error for many molecular simulation workloads~\cite{tranter2019ordering}

\paragraph{Lexicographic ordering.}
Terms are sorted alphabetically based on the corresponding Pauli string. This method has been shown to be useful in maximizing single-qubit gate cancellation chances across consecutive Pauli terms~\cite{gui2020term}. 

\paragraph{Gray-inspired ordering.}
This policy aims to place terms with similar supports adjacent in the schedule, so that the support changes smoothly and requires less qubit movement between consecutive gadgets. For each term $P$, define the \emph{Gray key}
\begin{equation}
  g(P) = M(P) \oplus \bigl(M(P) \gg 1\bigr),
  \label{eq:gray-key}
\end{equation}
where $\oplus$ is bitwise XOR and $\gg 1$ is a one-bit right shift. Terms are sorted by $g(P)$ rather than $M(P)$. Adjacent Gray-code values differ in exactly one bit, so this transform clusters support masks that differ in only a few qubits. We refer to this rule as \emph{Gray-inspired} rather than exact, because the sparse set of Pauli supports encountered in a Hamiltonian does not in general trace a true Gray-code path.

\paragraph{Schedule smoothness.}
To quantify the quality of an ordering, we use the \emph{support-delta surrogate}
\begin{equation}
  \label{eq:support_delta}
  \Delta_{\mathrm{sched}} = \sum_{t=2}^{T} d_H(M_{t-1}, M_t)
  = \sum_{t=2}^{T} \bigl|S_{t-1} \,\triangle\, S_t\bigr|,
\end{equation}
where $d_H$ is Hamming distance and $\triangle$ is symmetric difference. This equals the cumulative number of qubit-support changes between consecutive terms; lower values indicate smoother schedules and are expected to correlate with lower inter-core communication pressure. Finding the permutation that minimises $\Delta_{\mathrm{sched}}$ is equivalent to the Travelling Salesman Problem (NP-hard); all the comparison-based sort heuristics above run in $O(T\log T)$.

\subsection{Parity Tree Synthesis and Multi-core Allocation}
\label{sec:method_allocation}

The synthesis and allocation stage processes the support sequence $(S_1,\dots,S_T)$ produced by the scheduler. For each term $P_t$, implementing the Pauli gadget $e^{-i\theta P_t}$ requires a \emph{parity-reduction network}: a spanning CNOT tree over $S_t$ that accumulates the joint parity of the support qubits into a root qubit, followed by an $R_Z(2\theta)$ rotation and the reverse CNOT tree (Section~\ref{sec:pauli_scheduling}). Any spanning tree over $S_t$ yields a valid network, so tree topology is a degree of freedom the this stage exploits to align CNOT edges with the current qubit layout and vice-versa. Hardware is modelled as a graph of cores with fixed per-core capacity; communication cost equals the shortest-path distance $d(\cdot,\cdot)$ between cores. Synthesis and allocation of a Pauli term can result in multiple circuit slices. The algorithm proceeds by iterating over the Pauli terms, maintaining two growing index variables $t$ and $l$ for term and slice respectively. For each slice, we write $\ell_l$ for the layout at slice $l$ and $\ell_l(q)$ for the core hosting qubit $q$.

\subsubsection{Misplacement score}
For each term $t$, the allocator computes a misplacement score $\bar{d}_t(q)$ for each qubit $q$, measuring its discounted average core distance to future interaction partners:
\begin{equation}
  \bar{d}_t(q) =
  \sum_{\tau \in \mathcal{W}_t(q)}
  \gamma^{\tau-t}\,\delta_\tau(q),
  \label{eq:misplacement-score}
\end{equation}
where $\mathcal{W}_t(q) = \{\tau \in \{t,\dots,\min(t{+}W{-}1,T)\} : q \in S_\tau\}$ is the lookahead window of Pauli terms/supports in which $q$ participates, $W$ is the window size, $\gamma\in(0,1]$ is a decay factor, and
\begin{equation}
  \delta_\tau(q) =
  \frac{1}{|S_\tau|-1}
  \sum_{\substack{r \in S_\tau \\ r \neq q}}
  d\!\left(\ell_t(q),\,\ell_t(r)\right)
\end{equation}
is the average core distance from $q$ to its co-support qubits at term step $\tau$. A low misplacement score indicates a qubit that is well positioned relative to upcoming terms and should be kept in place; a high misplacement score marks a qubit that is a better candidate for displacement.

\subsubsection{Parity forest construction}
At this point, the parity CNOT tree has to be synthesized for the term $t$.
The allocator groups the support qubits $S_t$ by their current core assignments and forms a \emph{local parity chain} within each occupied core. Inside a core, the support qubits are chained into a directed CNOT sequence ordered by ascending misplacement score; the qubit with the highest misplacement score serves as the chain root. All CNOT edges within a chain are locally executable at zero communication cost. Each chain root is designated the core's \emph{representative}, and the collection of all per-core chains forms the initial \emph{parity forest}: a set of disjoint, locally executable CNOT sub-trees.

\subsubsection{Meeting core}
The disjoint subtrees on different cores need to be merged into a single parity tree. To this end, the allocator selects a \emph{meeting core} $c_t^\star$ as the core minimising the weighted distance to the support qubits, with qubits with lower misplacement score (better positioned) contributing more strongly:
\begin{equation}
  c_t^{\star} =
  \arg\min_{c \in \mathcal{A}(S_t)}
  \sum_{q \in S_t} \frac{d\!\left(c,\,\ell_t(q)\right)}{1 + \bar{d}_t(q)},
  \label{eq:gather-core}
\end{equation}
where $\mathcal{A}(S_t)$ is the set of active cores, i.e. cores containing at least a qubit from $S_t$ according to current layout. The meeting core acts as a spatial attractor toward which the parity tree is contracted.

\subsubsection{Forward merge phase}
Starting from the parity forest of $k$ disjoint sub-trees (one per active core), the forward phase iteratively merges them into a single spanning parity tree. At each iteration, the pair of representatives on the two closest cores is selected. One representative is moved to the other's core, generating a new circuit slice and adding core distance to the running inter-core communication count. The preferred merge direction is selected according to the score
\begin{equation}
  \Psi_t^{(1 \to 2)}
  = \Delta_{\mathrm{future}}^{(1 \to 2)}
  + \lambda \, \Delta_{\mathrm{meeting}}^{(1 \to 2)},
  \label{eq:merge-score}
\end{equation}
where $\Delta_{\mathrm{future}}^{(1\to 2)}$ is the discounted reduction in future interaction distance gained by moving the representative from core $c_1$ to $c_2$,
\begin{equation}
  \Delta_{\mathrm{meeting}}^{(1 \to 2)}
  = d(c_1,\,c_t^{\star}) - d(c_2,\,c_t^{\star}),
\end{equation}
and $\lambda \ge 0$ is a bias parameter, which we set to $\sfrac{1}{2}$. The direction with the larger score is preferred. If the destination core is full, the move is realised by swapping with the highest misplacement score (worst-positioned) resident qubit (excluding the core representative). If a swap is required, the communication cost count is increased by two times the core distance $d(c_1,c_2)$. After the two representatives are in the same core, the new root for the resulting subtree (representative of the core) is selected as the qubit with lowest misplacement score.

\subsubsection{Backward co-location phase}
The algorithm proceeds by merging subtrees iteratively until only a single tree remains.
Once all subtrees have been merged into a single parity tree, the tree topology for the Pauli gadget is fully determined. The backward phase traverses the tree from root to leaves, processing each parent-child edge in the reverse order of the parity-accumulation cascade to allow the execution of CNOT in the reverse parity tree of the Pauli gadget. For each edge where parent $p$ (on core $c_p$) and child $q$ (on core $c_q \neq c_p$) are not co-located, the allocator chooses the move direction by comparing the discounted future interaction distances:
\begin{align}
  G_t^{\mathrm{child}}  &= F_t(q,c_q) - \bigl[d(c_q,c_p) + F_t(q,c_p)\bigr], \label{eq:gain-child}  \\
  G_t^{\mathrm{parent}} &= F_t(p,c_p) - \bigl[d(c_p,c_q) + F_t(p,c_q)\bigr], \label{eq:gain-parent}
\end{align}
where
\begin{equation}
F_t(q,c) =
\sum_{\substack{\tau \in \mathcal{W}_t(q) \\ \tau > t}}
\gamma^{\tau-t}
\sum_{\substack{r \in S_{\tau} \\ r \neq q}}
d\!\left(c, \ell_t(r)\right)
  \label{eq:future-cost}
\end{equation}
is the discounted future interaction distance of placing qubit $q$ on core $c$. The qubit with the larger gain is moved; if the core is full the same swap strategy of forward pass is applied. Each co-location move generates a new circuit slice, updates the layout for subsequent edges in the same traversal and increases the running communication cost by $d(c_p,c_q)$.

After both phases complete, the allocator has produced a fully executable parity CNOT tree for $P_t$ together with the complete sequence of inter-core transfers required to realise it in both forward and reverse passes. The process repeats for $t = 1, \dots, T$, adding circuit slices as needed and choosing qubit layout for each of them.